\documentclass[a4paper,11pt]{article}
\pdfoutput=1
\usepackage{jinstpub}
\usepackage{graphicx,wrapfig}
\usepackage{breqn}
\usepackage{multirow}
\usepackage{lineno}
\usepackage{subfigure}
\usepackage{cancel}
\usepackage{gensymb}
\usepackage{xcolor}
\usepackage{array}
\usepackage{verbatim}
\newcolumntype{P}[1]{>{\centering\arraybackslash}p{#1}}
\newcolumntype{C}[1]{>{\centering\arraybackslash}m{#1}}
\begin{document}

\title{Performance of Hamamatsu VUV4 SiPMs for detecting liquid argon scintillation}


\author[a]{T. Pershing}
\author[a]{J. Xu } 
\author[a]{E. Bernard }
\author[a,b]{J. Kingston }
\author[a,c]{E. Mizrachi}
\author[a]{J. Brodsky} 
\author[e]{A. Razeto} 
\author[f]{P. Kachru} 
\author[a]{A. Bernstein}
\author[b] {E. Pantic}
\author[d]{I. Jovanovic}
\affiliation[a]{Lawrence Livermore National Laboratory, 7000 East Ave., Livermore, CA 94550, USA}
\affiliation[b]{University of California Davis, Department of Physics, One Shields Ave., Davis, CA 95616, USA}
\affiliation[c]{University of Maryland, Department of Physics, College Park, MD 20742}
\affiliation[d]{University of Michigan, 2355 Bonisteel Blvd, Ann Arbor, MI 48109, USA}
\affiliation[e]{INFN Laboratori Nazionali del Gran Sasso, Assergi (AQ) 67100, Italy}
\affiliation[f]{Gran Sasso Science Institute, L\'Aquila 67100, Italy}



\emailAdd{pershing1@llnl.gov}

\abstract{Detection of light signals is crucial to a wide range of particle detectors. In particular, efficient detection of vacuum ultraviolet (VUV) light will provide new opportunities for some novel detectors currently being developed, but is technically challenging. In this article,  
we characterized the performance of Hamamatsu VUV4 silicon photomultipliers (SiPMs) for detecting VUV argon scintillation light without wavelength shifting. 
Using a customized cryogenic amplifier design, we operated two models of VUV4 SiPMs inside liquid argon and thoroughly examined their direct sensitivities to liquid argon scintillation. In addition to describing their cryogenic performance, we measured a photon detection efficiency of $14.7^{+1.1}_{-2.4}$\% and $17.2^{+1.6}_{-3.0}$\%  
at 128 nm for these two VUV4 models for operation at 4 V of overvoltage, with the main uncertainty arising from the SiPM reflectivity for VUV light. 
}

\begin{NoHyper}
\maketitle
\end{NoHyper}


\section{Introduction}
\label{sec:intro}

In recent years, Silicon PhotoMultipliers (SiPMs) have become an attractive alternative to PhotoMultiplier Tubes (PMTs) for photodetection in a range of radiation detectors \cite{Sutanto2021,Cherry2011,ALEKSEEV2018,Darkside20kPDP,nEXOPDP,LEGEND1000PDP,PROTODUNESP,JUNO2020}.  SiPMs are capable of single photon detection with excellent resolution and can exhibit high photon detection efficiencies ($>50\%$ at 450nm) in the visible wavelength regime \cite{HamamatsuHighPDEBrochure}. 
SiPMs are composed of many micrometer-scale photosensitive cells and can, in principle, be produced in any planar geometry; most commercially available SiPMs are square, which tile well for filling surfaces with minimal dead space and maximizing photon detection coverage.   Relative to PMTs, their shallow depths also allow for reduced detector sizes, and their lower bias voltages relieve the need for high voltage power supplies. 
For experiments that require low radioactive background levels, SiPMs are also advantageous due to their lower radioactivity than that of standard PMT glass, enabling closer placement to sensitive detector volumes without an excessive increase in radioactive backgrounds \cite{SiPMRadioactivity}.  

However, SiPMs have their drawbacks to consider.  
Most notably, SiPMs have extremely high dark noise rates at room temperature (up to $\mathcal{O}$(MHz/mm$^2$)), which can limit their application in low-photon-statistics experiments. Fortunately, for low-temperature applications the dark noise rate in a SiPM decreases drastically, and dark rates below $\mathcal{O}$(Hz/mm$^2$) have been measured at liquid xenon and argon temperatures \cite{nEXOSiPM,Wang2021}. 
Another drawback is that SiPMs with satisfactory sensitivity to ultraviolet light, such as that emitted by noble element scintillators including argon and xenon, are still in the developing phase. 
Several large liquid argon and xenon experiments that search for rare processes, such as dark matter interactions and neutrinoless double beta decay, are actively investigating the use of large-area SiPM modules for light detection to enhance the experiments' sensitivity  \cite{Falcone2020,DIncecco2017}. To study the feasibility of such SiPM applications, the performance of newly developed SiPMs needs to be thoroughly characterized in noble element detectors. For liquid argon detectors, due to the extremely short scintillation wavelength (128 nm) and the lack of photosensors with adequate sensitivity to this wavelength region, 
virtually all experiments still require the application of wavelength shifters such as 1,1,4,4-tetraphenyl-1,3-butadiene (TPB) to detect the scintillation light \cite{WLSSurvey,DEAPTPB}.

Hamamatsu has recently developed the VUV4-series of SiPMs, which report a Photon Detection Efficiency (PDE) comparable to that of PMTs at the xenon wavelength of 174 nm, and also appreciable PDEs down to the peak VUV argon scintillation wavelength of 128 nm \cite{HamamatsuVUV4Brochure}.   
This significant improvement of VUV performance over models that were previously available 
provides a real opportunity for these SiPMs to be deployed in future noble element detectors.
In this work, we characterize the performance of VUV4 SiPMs for the detection of liquid argon scintillation light; measurements are made in a configuration resembling a typical argon detector setup to evaluate their suitability for deployment in such experiments.
A rigorous examination of the direct VUV sensitivity of VUV4 SiPMs is presented through analysis of the observed signals from SiPMs with different wavelength sensitivities.  Measurements of the gain, dark noise, cross talk and afterpulsing as a function of overvoltage in liquid argon are also shown. Finally, a PDE estimate for the Hamamatsu VUV4 SiPMs is made at the argon scintillation wavelength of 128 nm in a liquid argon environment and compared to the quoted PDEs from Hamamatsu and values previously measured for the earlier VUV3 models~\cite{Igarashi2016_VUV3}.

\section{Experimental setup}
\label{sec:detector}

\subsection{Hamamatsu VUV4 SiPMs}

Two packages of Hamamatsu VUV4 SiPMs were evaluated in this work. 
Both packages  are potential candidates for a future deployment in  
small xenon and argon detectors being developed at Lawrence Livermore National Laboratory.  The first package is the S13371-6050CQ-02, containing four SiPM cells mounted in a single compact ceramic package.  
The second package is the S13370-6075CN, housing a single SiPM cell mounted in a ceramic package. 
The key properties of these SiPMs as measured by Hamamatsu are provided in Table \ref{SiPMPropertyTable}.  
\begin{table*}[ht]
\centering
\caption{Key properties of the S13371 and S13370 models of Hamamatsu's VUV4 SiPMs.  The S13371 contains four SiPMs per package; properties quoted are for each individual SiPM in the package.  Properties are characterized at a temperature of 25\degree C and an overvoltage of 4 V.  Further details available at \cite{HamamatsuVUV4Brochure}.}

\vspace{0.3cm}

\footnotesize
\setlength\tabcolsep{2.5pt} 
\begin{tabular}{|C{0.11\linewidth} | C{0.10\linewidth} | C{0.10\linewidth} |C{0.10\linewidth}|C{0.10\linewidth}|C{0.09\linewidth}|C{0.10\linewidth}|C{0.08\linewidth}|C{0.09\linewidth}|}
\hline
Model & Breakdown voltage (V)  & Gain (unitless) & Package size (mm$^2$) & Active area (mm$^2$) & Crosstalk prob. (\%) & pixel pitch ($\mu$m) & Terminal cap. (nF) & Fill factor within cell perimeter (\%)\\
\hline
\hline
S13371-6050CQ-02&$53\pm5$ & $2.55 \times 10^{6}$ & $15 \times 15$ & $6 \times 6$ (each of 4 cells) & $3$ & 50 & 1.20 (each of 4 cells) & 60\\
\hline
S13370-6075CN & $53\pm5$ & $5.8 \times 10^{6}$ & $10 \times 9$ & $6 \times 6$ &  $5$ & 75 & 1.28 & 70\\
\hline
\end{tabular}
\label{SiPMPropertyTable}
\end{table*}

A key difference between the two packages is that S13371-6050CQ-02 has a quartz window in front of the sensitive SiPM area, while the S13370-6075CN does not. Optical grade quartz has a transmission cutoff near 170 nm and thus renders the S13371-6050CQ-02 units ineffective in detecting 128 nm VUV light from argon scintillation \cite{Newport}. In this work, one S13371-6050CQ-02 unit had its quartz window removed to allow 128 nm photons to reach the sensitive SiPM surfaces. 
In contrast, the S13370 model does not have a quartz window and is rated for 128 nm light detection by the manufacturer. However, it has a large insensitive packaging frame around the sensitive SiPM area, which can result in a lower effective photosensitive coverage compared to the S13371-6050CQ-02 model.

\subsection{Cryogenic preamplifier circuit}
In photodetection applications requiring single-photon sensitivity, one challenge is that SiPM pulses typically have widths of tens to hundreds of nanoseconds due to a long pulse tail (versus <10 ns pulse widths typical for PMTs) from increased capacitance.  The result of this is a smaller single photoelectron peak amplitude than PMTs at similar gains ($\mathcal{O}$($1 \times 10^{6} $)), which can result in lower trigger efficiency and larger deviation in measured pulse integrals due to baseline noise.  Compensating for this by operating the SiPMs at high overvoltages to achieve high gains may degrade the single photoelectron charge resolution through increased afterpulsing and crosstalk.  Optimal performance can be obtained by pairing SiPMs with low-noise amplifiers to achieve $\mathcal{O}$($1 \times 10^{7} $) gain at relatively low overvoltages that do not lead to excessive crosstalk and afterpulsing \cite{DIncecco2018,Carniti2020,Falcone2021}, yet enable efficient detection of SPE pulses at the mV-scale.

This work utilizes the LMH6629 operational amplifier from Texas Instruments to provide the gain needed for efficient detection of single photoelectron pulses with low crosstalk and afterpulsing.  The LMH6629 
utilizes Heterojunction Bipolar Transistors (HBTs) made of SiGe; the performance
of SiGe HBTs degrades less at cryogenic temperatures than that of traditional Silicon
Bipolar Junction Transistors (BJTs) \cite{1481524}, making them a good preamplifier choice for
operation in liquid argon.  Additionally, examples of successful 
SiPM preamplification with the LMH6629 in the WSON-8 package described in the literature are useful for cross-checking performance and informing circuit design.  

\begin{figure}[!ht]
    \centering
    \includegraphics[width=1.0\textwidth]{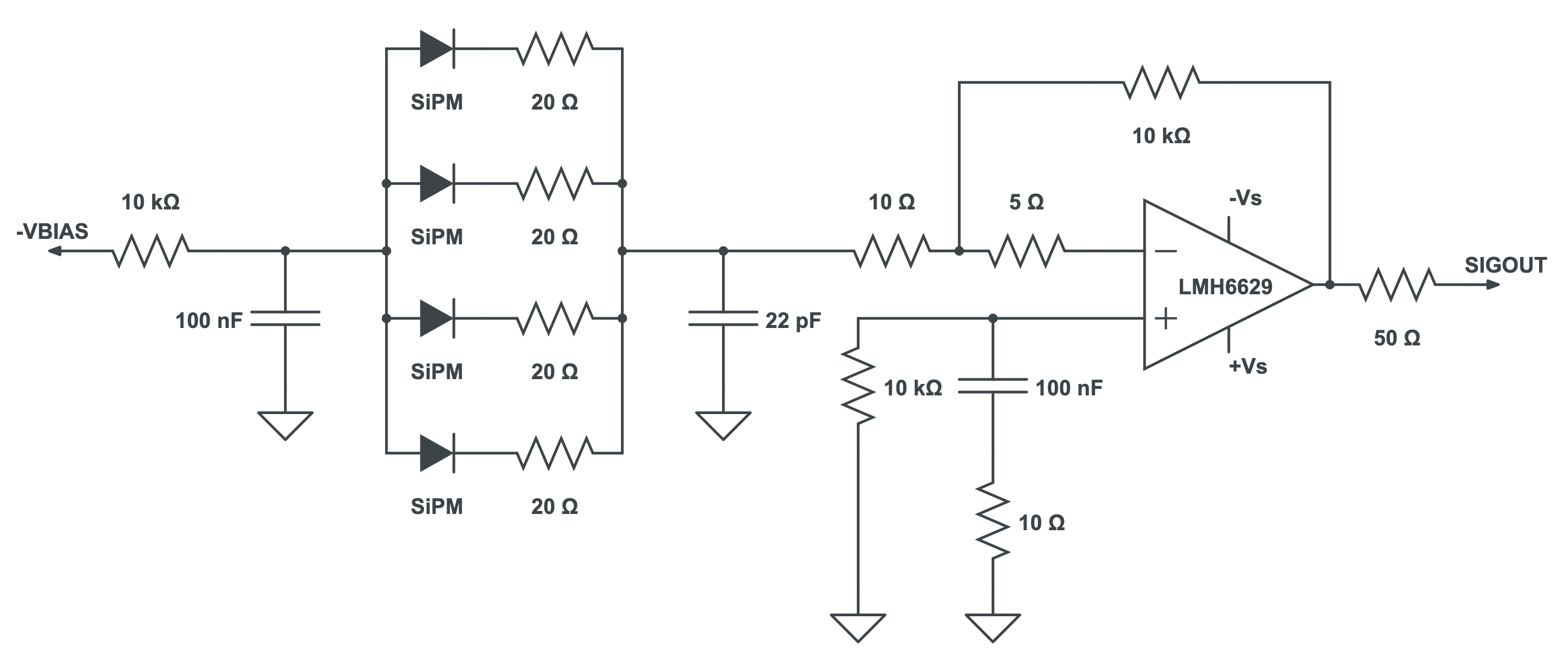}
    \caption{ Circuit diagram of the SiPM biasing scheme and LMH6629 readout.}
    \label{fig:AmpCircuit}
\end{figure}

The cryogenic TransImpedance Amplifier (TIA) circuit used with the Hamamatsu VUV4 SiPMs is shown in Fig.\ref{fig:AmpCircuit}. The circuit design is a modified version of that described and characterized in \cite{DIncecco2017}. The
first capacitor and resistor form a low-pass filter for providing a constant, low-noise bias voltage to the SiPMs.
Four SiPM cells (either from a single S13371 unit, or four S13370 units) are placed in parallel and read into a 
single LMH6629.  The resistors at each SiPM's output reduce the current flow from an active SiPM backward into the 
inactive SiPMs.  The feedback resistor value can be adjusted to control the gain at the output, but
may also affect the stability of the TIA amplifier circuit \cite{AnalogArticle}.
To minimize variation in passive component values at cryogenic temperatures, all capacitors are of NPO C0G type 
and all resistors are either thick or thin metal film resistors.

For this measurement, all amplifiers were operated with a supply voltage of $\pm 1.45 $ V, for a total supply voltage
of 2.9 V.  Higher supply voltages at liquid argon temperatures were 
observed to induce oscillations at the signal amplifier output 
due to the increase in the op-amp's gain-bandwidth product \cite{DIncecco2018}.  
While the present circuit design is deemed sufficient for the SiPM characterization reported in this work, modifications are being pursued to allow for operation with higher amplifier supply voltages.

\subsection{Liquid argon detector for SiPM studies}
\label{subsec:cryostat}
\begin{figure}[!ht]
\centering
\begin{minipage}{.54\textwidth}
  \centering
  \includegraphics[width=1.0\linewidth]{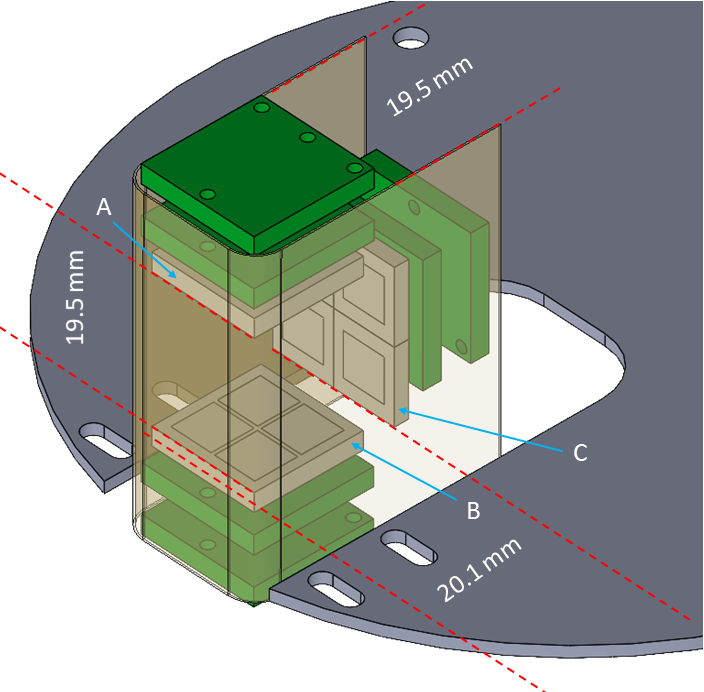}
\end{minipage}%
\hspace{0.4cm}
\begin{minipage}{.40\textwidth}
  \centering
  \includegraphics[width=1.0\linewidth]{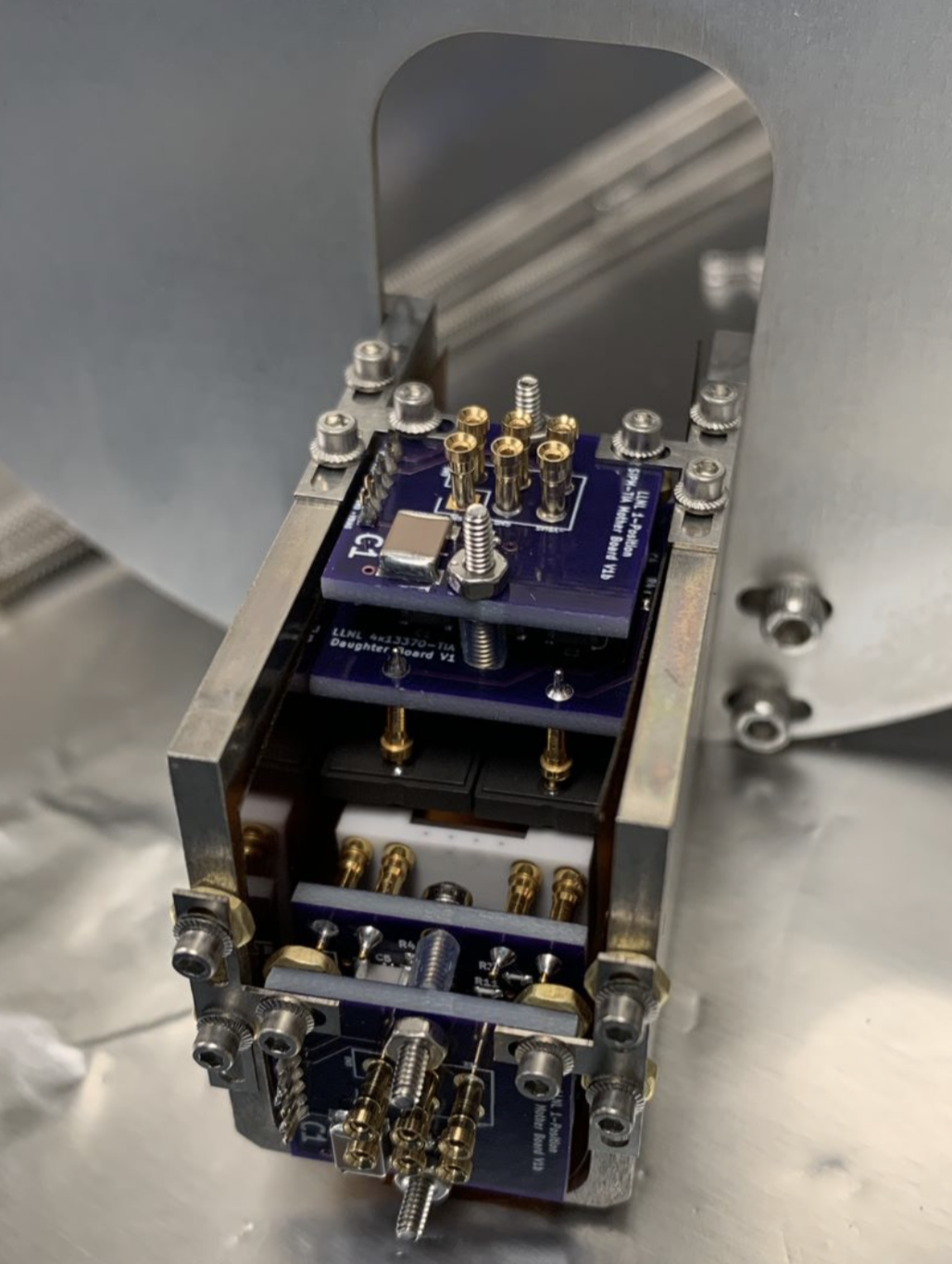}
\end{minipage}
\caption{Left: CAD rendering of SiPM cell design.  The aluminum walls that support the Kapton sheet are omitted for clarity.  A: S13371 module with quartz window.  B: S13371 module without quartz window.  C:  Array of four windowless S13370 modules.  Right: Photo of mounted SiPM-TIA boards.  A two-board design is utilized, one for mounting the SiPM and amplifier and another for providing power and signal readout pin headers.  All three SiPM-TIA units are immersed in liquid argon during operation.
}
\label{fig:SiPMCell}
\end{figure}

Three SiPM-TIA modules, the first being a S13371 module with a quartz window, the second a S13371 module with the quartz window removed, and the third a S13370 module containing four windowless S13370 units in a 2x2 array, were deployed for this study. As illustrated in Fig.~\ref{fig:SiPMCell}, the three SiPM modules were mounted around a cuboid volume of liquid argon of 19.5 mm x 19.5 mm x 20.1 mm dimensions.  The other three faces of the argon target were enclosed by a Kapton sheet of 0.25 mm thickness to define the boundaries of the active volume.
This simple detector cell is submerged within and positioned near the wall of a larger liquid argon bath.  During operation, the argon liquid level was maintained at $\sim$ 4 cm above the top SiPM window, fully submerging all SiPM-TIA modules. 

The detector and liquid argon bath are placed in a vacuum-insulated cryostat and are cooled by a thermosiphon loop with argon as the working fluid.  The thermosiphon loop is cooled by a Sumitomo Cryogenics CH-104 Gifford-McMahon helium cycle cryocooler.  Argon within the detector volume was commercially procured at 99.9997\% purity in a high-pressure cylinder and further purified through a SAES Mono-Torr PF3-C3-R-1 hot metal getter before it was condensed and delivered to the detector volume. No continuous purification of the argon was used in this experiment, but thanks to the low impurity outgassing rate from detector components at liquid argon temperature, no significant degradation of argon scintillation was observed over the operation period of 12 days. For all data acquired, the liquid argon temperature was maintained at 91.2 K through an active PID loop, yielding a system pressure of 1.5 bar.

Bias voltages to the three SiPM units  were individually controlled
using a CAEN R8031N 8-channel power supply (0.03 \% long-term stability, $<$ 5 mV ripple).  All three TIAs were powered with a single Teledyne T3PS23203P power supply ($<$ 0.35 mV ripple).  Signals from each SiPM module were split using a CAEN N625 linear fan-out, with one copy sent to a CAEN N844 low-level discriminator to generate triggers and the other copy sent to the inputs of Struck SIS3316 digitizers (14-bit depth over 2 V input range, 4 ns per sample) to be recorded. Unless otherwise specified, a threshold-crossing in either the windowless S13371 SiPM-TIA module or the 4xS13370 SiPM-TIA module triggered data acquisition from all three SiPMs. For each trigger, 12 $\mu{\mathrm s}$ of waveform was recorded for each SiPM, starting from -2 $\mu{\mathrm s}$ before the trigger time.

The CAEN N844 discriminator levels were configured to yield 100\% trigger efficiency within uncertainties for each SiPM module at the lowest overvoltages studied in this work.  To confirm this, data was taken from all SiPM modules along with a full digitization of the discriminator pulses.  All pulses measured in each SiPM module 5 $\mu$s after the initial trigger were cross-checked against their respective discriminator channel to determine the probability of triggering the discriminator as a function of pulse amplitude.  

After the raw SiPM waveforms were stored on a computer disk, the data was read into offline analysis software, where waveform baselines were calculated and subtracted, and pulses down to single photoelectrons were identified. For every pulse in each SiPM channel, quantities including pulse area and amplitude were calculated and stored into a ROOT ntuple format. The pulse-processing was an adapted version of the DAQMAN software package, which is open-source and can be found at \cite{DAQMAN}.

\section{Verification of 128 nm light detection} 

To produce liquid argon scintillation light of known intensity, an Am-241 source was placed outside the cryostat vacuum can adjacent to the SiPM-instrumented active liquid argon volume.  Am-241 undergoes alpha decay producing an accompanying 59.5 keV gamma 35.9\% of the time. 
The mean free path of 59.5 keV gammas is estimated to be $\sim$0.1 cm in stainless steel and $\sim$1.5 cm in liquid argon. Therefore, a useful fraction of Am-241 gammas can reach the liquid argon and deposit all energy in the instrumented active volume.

\begin{figure}[!ht]
\centering
  \centering
  \includegraphics[width=0.9\linewidth]{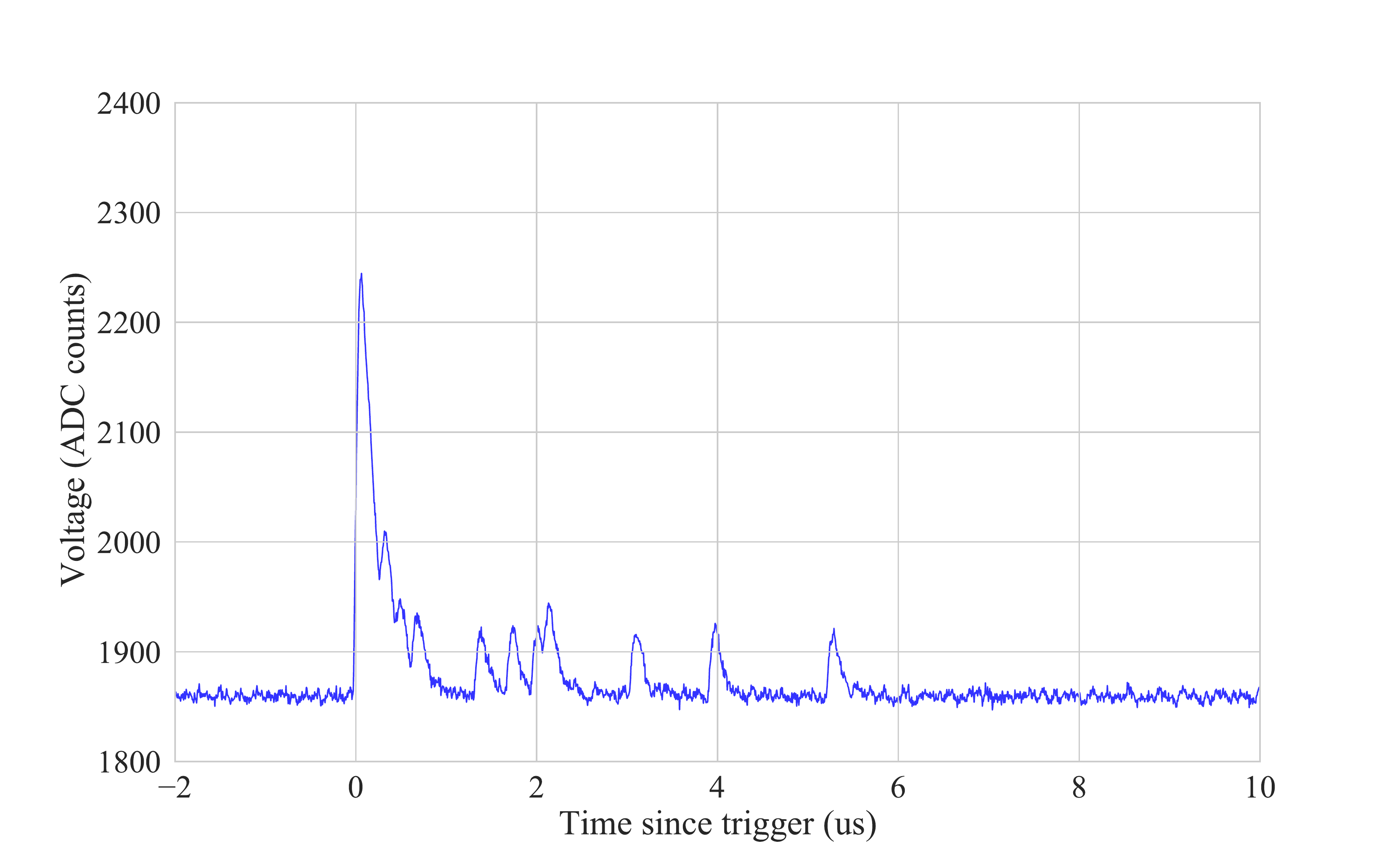}
\caption{Single acquisition of an Am-241 candidate event in the windowless S13371 unit at $1.97 \, {\mathrm V}$ overvoltage.  A prompt pulse is seen at the trigger time, followed by single photoelectron pulses.}
\label{fig:Waveforms}
\end{figure}

Recoiling electrons produced by Am-241 gamma interactions can generate a track of free electrons, positive argon ions, and excited atomic argon atoms.  Molecular argon excimer dimers are formed through the pairing of excited argon atoms with neighboring ground state argon atoms and by the recombination of electrons with ionized argon~\cite{Hitachi1983}.  The dimers are metastable and are produced in singlet and triplet states, which decay to produce 128 nm VUV scintillation light at fast and slow timescales, respectively.  An example SiPM waveform of liquid argon scintillation measured in this work is shown in Fig.\ref{fig:Waveforms}, which consists of a prompt peak (at time 0) and a tail of sparse photoelectron peaks persisting for several microseconds; these features are consistent with the expected fast and slow decay times, as shown in table \ref{tab:TripletDecay}.

However, unintended wavelength shifting of argon scintillation to longer wavelengths by fluorescent detector components around the active liquid argon and the SiPMs could produce similar observations, even if the photosensors had no direct VUV sensitivity, and this possibility requires investigation. 
In this section, we first examine the recorded SiPM waveforms to confirm their origin from argon scintillation, 
and then we compare signals from  SiPM modules with different wavelength sensitivities to verify that these signals result from the direct detection of 128 nm light.

\subsection{SiPM waveform studies}

\begin{figure}[!ht]
\centering
\begin{minipage}{.49\textwidth}
  \centering
  \includegraphics[width=1.0\linewidth]{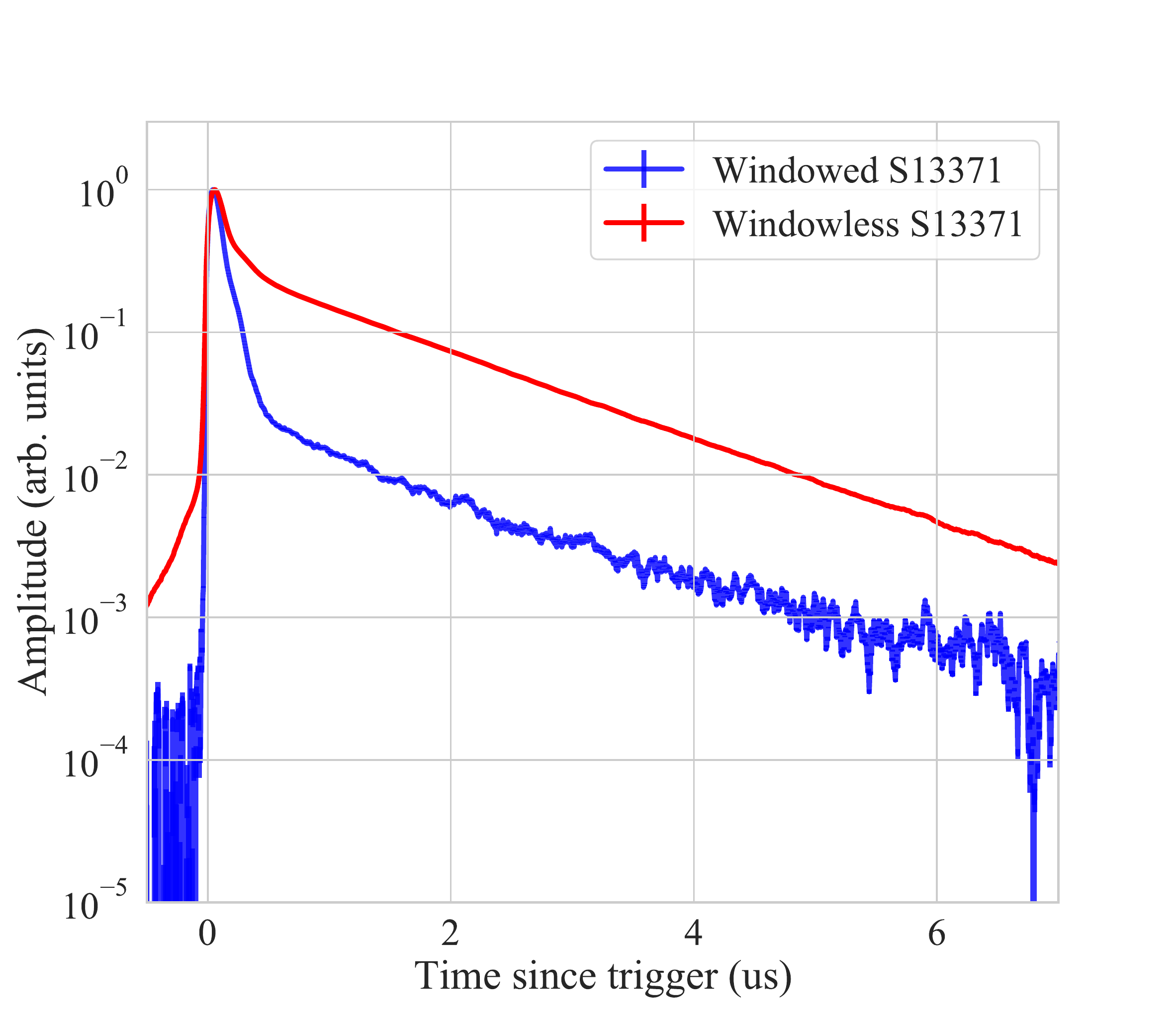}
\end{minipage}
\begin{minipage}{.49\textwidth}
  \centering
  \includegraphics[width=0.96\linewidth]{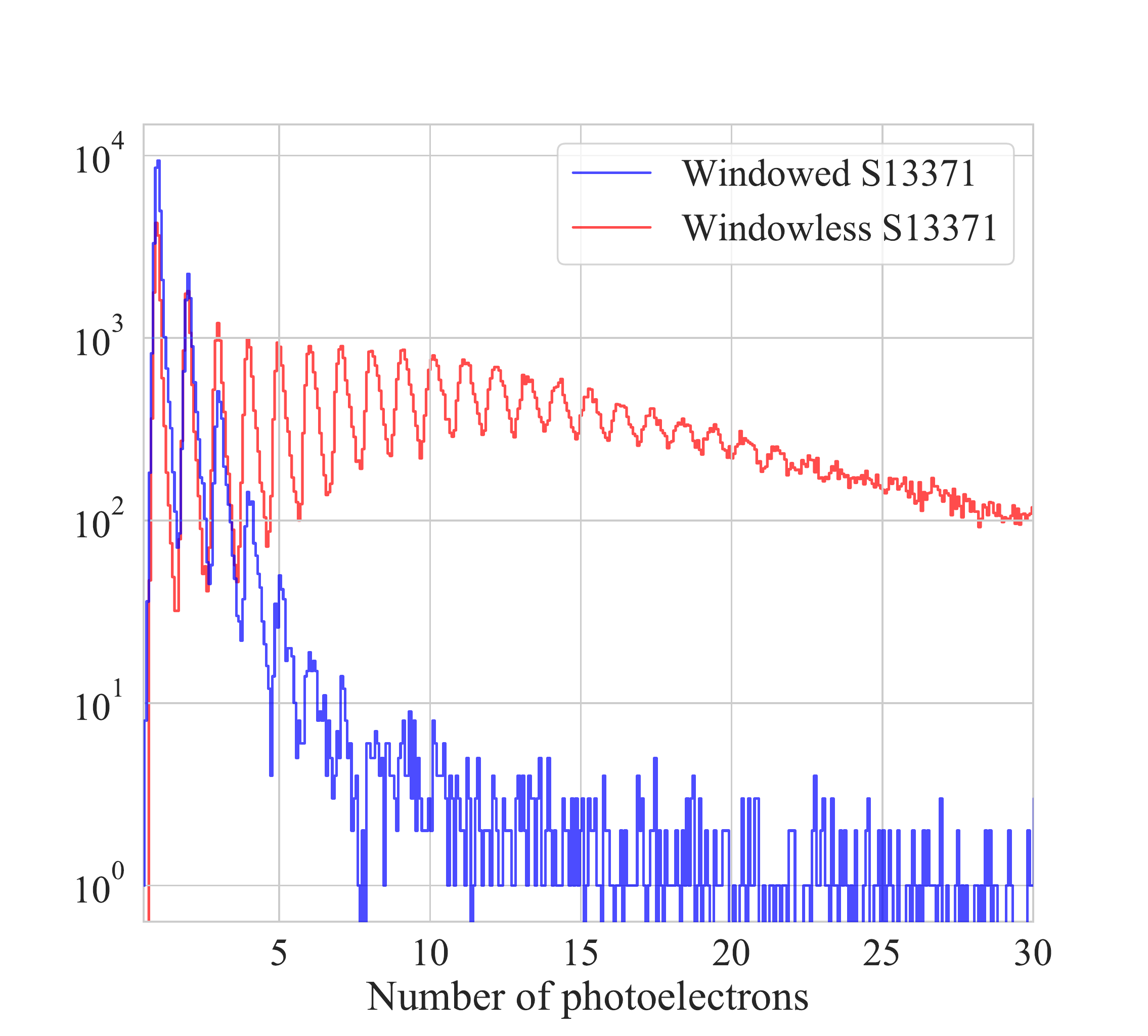}
\end{minipage}%
\caption{Left: Average waveforms observed in the windowed S13371 SiPM (blue) and in the windowless S13371 SiPM (red) in a single Am-241 calibration run. The windowed SiPM average waveform has a larger prompt scintillation component than the windowless SiPM. Right: Measured PE distribution for the windowed (blue) and windowless (red) S13371 during an Am-241 calibration run.  The windowed S13371 measures only a few photoelectrons on average per event, while the windowless S13371 measured tens of photoelectrons.}
\label{fig:WWLComp}
\end{figure}

Figure~\ref{fig:WWLComp} (left) shows the average waveform for thousands of Am-241 signals in the windowless S13371 module in red. Waveforms containing less than three photoelectrons in the windowless S13371 signal were excluded from this average to remove contamination from dark noise.  
The triplet state de-excitation time constant is estimated by fitting an exponential function to the average waveform for the windowless S13371 and S13370 units following the prompt peak.   
The exponential fit was performed in the time window from between 1.5 and 7 microseconds following the trigger time, excluding the prompt peak that is dominated by the singlet de-excitation. The fitted triplet de-excitation time is compared to typical measured values in the literature in Table \ref{tab:TripletDecay}, where our reported value is 
the mean of the two best fit values in each SiPM, with the uncertainty taken as half of the difference of the two measured values.  A systematic uncertainty is also quantified in each fit by translating the fit window 0.5 $\mu$s earlier and later across the waveform and taking the difference with the original fit; this uncertainty is added in quadrature with the difference of the measured values.  

\begin{table*}[ht]
\centering
\caption{Triplet state de-excitation time $\tau_T$ and the singlet-to-triplet intensity ratio $I_{s}/I_{T}$ measured in this work compared to other past measurements.  Reference values adapted from \cite{Hitachi1983}.}

\vspace{0.2cm}

\footnotesize
\begin{tabular}{|c|c|c|c|}
\hline
Measurement & This work & Hitachi & Carvalho and Klein \\
\hline
\hline
$\tau_{T}$ (ns)   & 1447 $\pm$ 20 & 1590 $\pm$100 & 1540 \\
\hline
$I_{s}$/$I_{T}$   & 0.29 & 0.3 &  0.26 \\
\hline
\end{tabular}
\label{tab:TripletDecay}
\end{table*}

Table~\ref{tab:TripletDecay} also compares the singlet-to-triplet ratio measured in this work to that reported in the literature.
The singlet component of the average waveform was calculated by extrapolating the triplet de-excitation exponential fit to a broader average waveform window of  0 to 8 $\mu$s.  The extrapolated triplet integral was then subtracted from the full waveform integral to yield the singlet component.  The singlet-to-triplet ratio was then calculated by dividing the integrated singlet component by the extrapolated triplet state exponential integral.

Both the triplet state de-excitation time and the singlet-to-triplet ratio measured in this work are in agreement with the values measured in different liquid argon detectors, confirming that signals from the Hamamatsu VUV4 SiPMs are indeed from liquid argon scintillation light.    

\subsection{VUV origin of SiPM signals}

A series of precautions were taken to mitigate possible contamination of non-VUV light in the SiPM signals. First, the active liquid argon target and three SiPM modules were grouped compactly and placed at the very edge of the liquid argon bath, as explained in Sec.\ref{subsec:cryostat}. This configuration minimizes the amount of auxiliary detector components, which might shift argon scintillation light to longer wavelengths~\footnote{We comment that no materials with known wavelength shifting properties were used in the construction of the liquid argon detector employed in this work, and no such materials were introduced to the system in past operations.}, exposed to the active argon and the SiPMs. Second, a Kapton film was used to surround the active argon volume on all sides not facing SiPMs. Kapton polyimide is known to strongly absorb UV light~\cite{French2009_KaptonOptical}, and its implementation in this experiment further reduces the probability for argon scintillation light in the target cell to reach auxiliary detector components outside the volume, possibly get wavelength-shifted, and then become detected by the SiPMs. In addition, the Kapton film also blocks the leakage of scintillation light from argon outside the instrumented volume and reduces possible reflection of VUV light from surrounding surfaces, and thus improves the robustness of optical modeling.  
Most importantly, a windowed S13371 unit and a windowless S13371 unit are mounted symmetrically around the active liquid argon volume, leading to approximately equal probability for an argon scintillation photon to reach either SiPM. Because quartz absorbs nearly all light below $170$ nm, the 128 nm liquid argon scintillation photons can only be efficiently detected by the windowless S13371 unit but not by the windowed S13371 unit. As a result, signals recorded by the windowed S13371 SiPM provides an \textit{in situ} measurement of the long wavelength photon background seen by the SiPMs.

Figure~\ref{fig:WWLComp} (right) shows the photoelectron spectra measured by the windowless (red) and windowed (blue) S13371 SiPMs under Am-241 radiation. For every Am-241 induced argon scintillation in the active volume, the SiPM without a quartz window typically detects tens of photoelectrons, while the S13371 SiPM with a quartz window only observes a few photoelectrons most of the time. Integrating both distributions, we find that the windowed SiPM measures only 1.5\% the amount of light that the windowless SiPM does. This result confirms that the windowless VUV4 SiPMs used in this experiment are predominantly detecting light below the $170$ nm quartz cutoff wavelength, specifically the $128$ nm component of argon scintillation light, 
and that contamination from longer wavelength photon backgrounds, which may result from unintended wavelength shifting processes in the detector, is negligible. 

The result of the experiment discussed above confirmed the effectiveness of the strategies used in this work to mitigate non-VUV photon contamination. 
However, because the Kapton film comprised the largest optical surface seen by the active liquid argon and SiPMs and the optical property of Kapton under strong VUV radiation is not yet fully understood, 
we conducted a follow-up experiment to investigate whether the use of Kapton could have biased the result. In this test, we removed the Kapton sheet surrounding the active xenon volume and repeated the measurement with Am-241. The observed results were consistent with that reported in Figure~\ref{fig:WWLComp} (right). Therefore, we report that polyimide does not produce significant fast fluorescence under VUV radiation, and this work unambiguously confirmed the direct VUV sensitivity of Hamamatsu VUV4 SiPMs in a liquid argon environment. 

Figure~\ref{fig:WWLComp} (left) also shows the average waveform recorded by the windowed S13371 SiPM with a single photoelectron threshold in blue~\footnote{In Am-241 calibration runs, the DAQ was triggered by the windowless SiPMs so the use of a single photoelectron threshold in the windowed SiPM didn't bias the result by including dark counts.}. In addition to observing only a small amount of light, the windowed S13371 also recorded signals that feature a dominant prompt component with a much suppressed delayed tail. This observation suggests that the prompt argon scintillation may include a component with a longer wavelength than that of the delayed triplet de-excitation light (128 nm), possibly in the longer wavelength UV as reported for gaseous argon in \cite{Santorelli:2020fxn}. 
The total triplet state light measured by the windowed SiPM is less than 1\% of that measured by the windowless SiPM, and may have originated from VUV light leakage into the SiPM around the quartz window, gamma interactions in the liquid argon between the quartz window and the SiPM sensitive face, optical crosstalk from the windowless SiPM units (photons produced by an active SiPM during avalanches leaving the SiPM \cite{Boulay:2022rgb}), or some unknown wavelength shifting process in the detector.

\section{SiPM characteristic measurements in liquid argon}

\begin{figure}[!ht]
    \centering
    \includegraphics[width=0.9\textwidth]{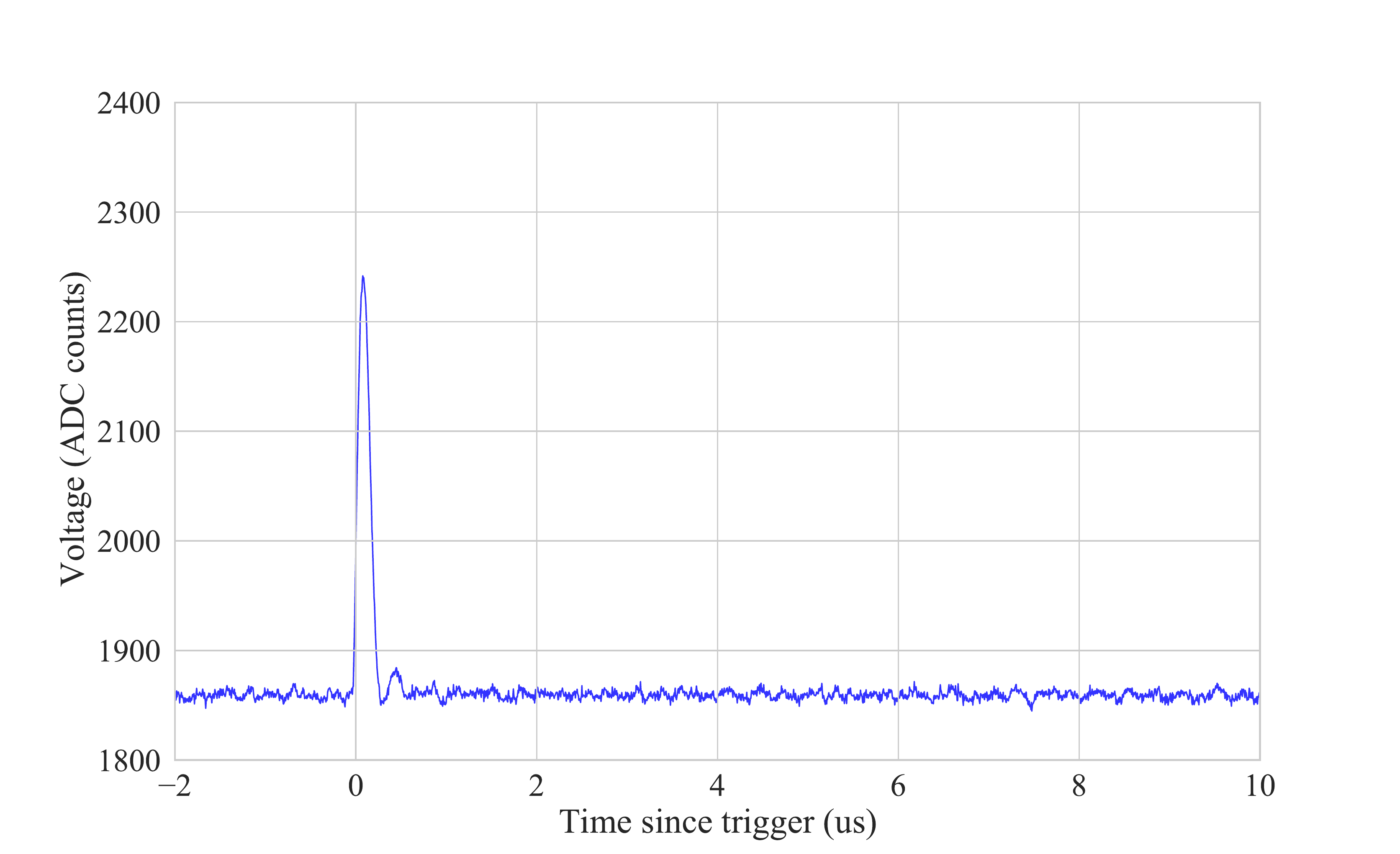}
    \caption{ A representative waveform in the windowless S13371 module from an Am-241 calibration run after oxygen quenching.  No triplet de-excitation pulses are visible following the prompt scintillation peak.}
    \label{fig:SingleWaveformOQ}
\end{figure}

To thoroughly characterize the performance of the S13371 and S13370 SiPMs in a liquid argon environment, we also acquired data with deliberately suppressed liquid argon scintillation.  The scintillation quenching was achieved by introducing a small amount of dry air into the detector volume.  Dissolved oxygen provides a rapid non-radiative de-excitation channel of Ar-Ar dimers that directly competes with the singlet and triplet decay processes that produce VUV scintillation light \cite{WArP2008}.  A typical SiPM waveform in the oxygen quenched data is shown in Fig.~\ref{fig:SingleWaveformOQ}, where no significant scintillation tail can be observed. For oxygen-quenched data-taking, any SiPM crossing a triggering discriminator threshold would begin data acquisition for all three SiPMs.  Seven hours of oxygen-quenched data were acquired with no radiation source present, and this data was used to evaluate the gain, dark noise, and afterpulsing probability of SiPMs.  For the study of  
crosstalk probability, which is prone to bias from simultaneous photon emission backgrounds, we also used SPE-like pulses in the late tail of Am-241-induced argon scintillation data prior to oxygen-doping. Given the known scintillation strength, the probability of simultaneous photoelectron detection in this time window is deemed negligible.

\subsection{Gain vs. voltage and breakdown voltages}\label{subsec:Gain}
To estimate the single photoelectron gain for all SiPM-TIA units, a Gaussian fit is performed on single photoelectron pulse area distributions in the oxygen-quenched argon data.  The range of the first fits to the mean are chosen manually, from approximately $2\sigma$ below the statistical SPE mean to 1.5$\sigma$ above the statistical mean.  This asymmetric fit range was chosen to mitigate bias from the afterpulsing effect (see section \ref{subsec:CTAP}), which increases the pulse area for SPEs when fitting too far to the right of the SPE peak. In addition,
a systematic uncertainty is included by refitting the Gaussian on the higher-charge distribution to only 0.5$\sigma$ past the initial SPE mean fit and to 2$\sigma$ past the initial mean fit. For all bias voltages, the total uncertainty of the SPE fit (fit uncertainty and systematic uncertainty added in quadrature) is small, less than 1\% relative to the SPE mean for all bias voltages studied in this work. 

\begin{figure}[!ht]
    \centering
    \includegraphics[width=0.85\textwidth]{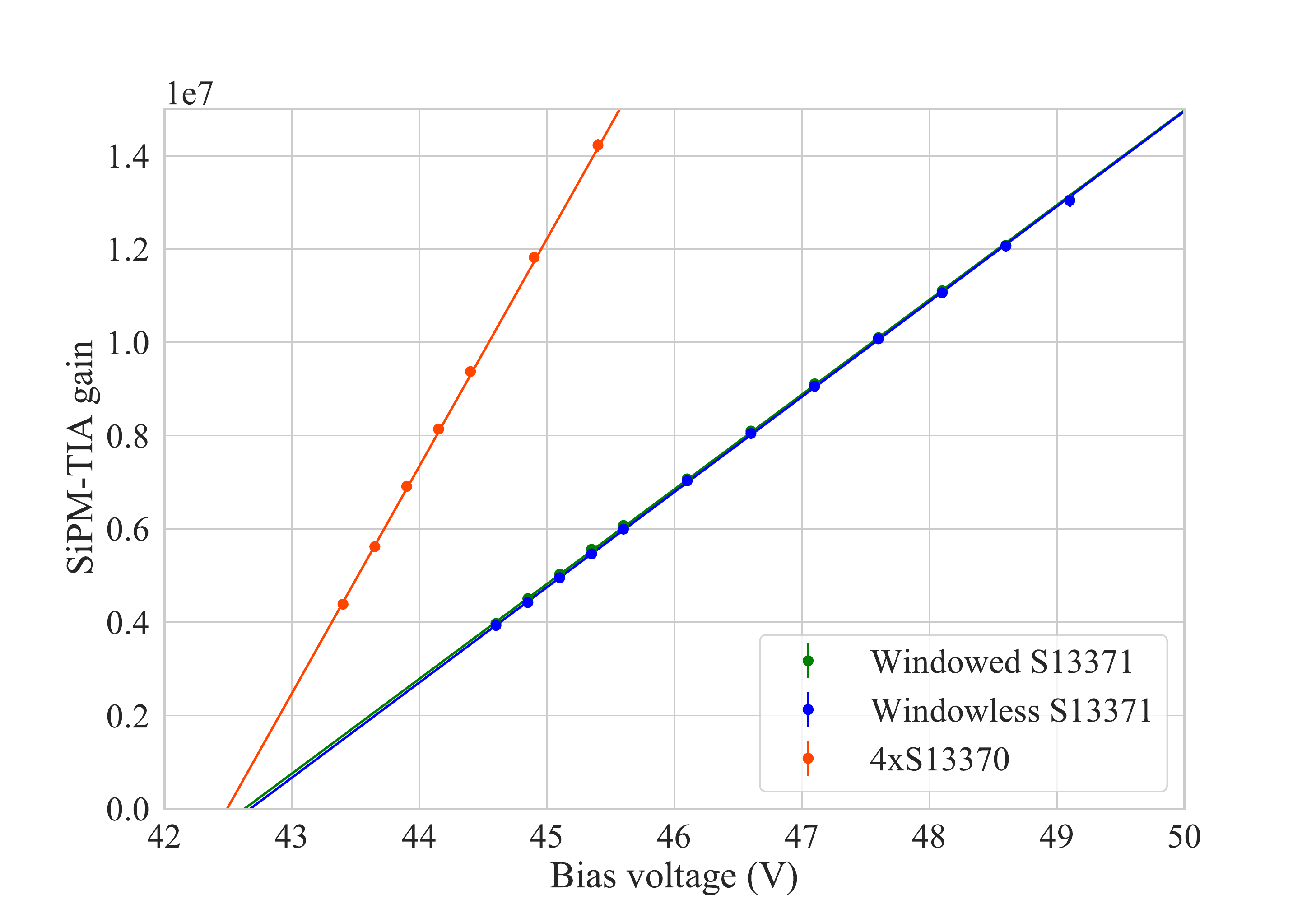}
    \caption{ Single photoelectron gains as a function of bias voltage for each SiPM-TIA module.  Both S13371 SiPM-TIA units are found to be in very close agreement regarding gain characteristics.  Lines show the best linear fit to gain curves for each module.}
    \label{fig:GainVsVoltage}
\end{figure}

After finding the single photoelectron integral mean in ADC counts, the SiPM-TIA single photoelectron gain is estimated by converting to total number of electrons measured at the digitizer. Best fits to the single photoelectron gain as a function of bias voltage for each SiPM-TIA module are shown in Fig.~\ref{fig:GainVsVoltage}.
The gain as a function of bias voltage is found to be linear for all three SiPMs, as expected.  The slope of the S13370 is more than twice that of the S13371, leading to higher gains at the same overvoltage.  This is in agreement with the relative gains quoted from Hamamatsu shown in Table \ref{SiPMPropertyTable}, where the S13370 has 2.27 times the gain of the S13371 at room temperature.  

The amplification factor of the TIA is estimated by comparing the SiPM-TIA gain vs.$\,$voltage curve
to the gain vs.$\,$voltage of a S13371 SiPM unit measured without the TIA in a separate measurement, and is found to be $76\pm1$.  This is lower than the gain of the TIA measured in \cite{DIncecco2017}, which may be due to operating the TIA in this work at a smaller supply voltage and consequently a smaller gain-bandwidth product.

The breakdown voltage for each SiPM unit can be estimated by fitting a straight line to each SiPM-TIA module's gain vs. bias voltage curve and finding the bias voltage value corresponding to a gain of zero.  The results for each SiPM's breakdown voltage estimate in liquid argon are shown in Table \ref{tab:Breakdown}. The variation in each SiPM model's breakdown voltage is expected, as different SiPM models (and individual SiPMs of the same model) can have different breakdown voltages.

\begin{table*}[ht]
\centering
\caption{Best fit breakdown voltages for all VUV4 SiPMs used in argon teststand.  Estimates were made with SiPMs immersed in liquid argon at 91.2 K and 1.5 bar.}

\vspace{0.2cm}

\footnotesize
\begin{tabular}{|c|c|}
\hline
Model & Breakdown voltage (v) \\
\hline
\hline
Windowed S13371 & 42.63 \\
\hline
Windowless S13371 & 42.67 \\
\hline
4xS13370 & 42.49 \\
\hline
\end{tabular}
\label{tab:Breakdown}
\end{table*}

\subsection{Dark noise rate limits}

In the absence of incident photons, SiPMs may also produce photoelectron pulses from thermally-generated carriers in a SiPM microcell's depletion region. These pulses are called dark noise, and are observed at up to $O$(MHz/mm$^2$) level at room temperature but are greatly reduced at cryogenic conditions. Due to the extremely low dark noise rate anticipated at liquid argon temperature and the presence of stray scintillation light in the oxygen-quenched argon from ambient radiation, we only report an upper limit on the rate in this work.
The upper bound on dark count rates as a function of overvoltage for each SiPM array is characterized by counting the total number of pulses observed in the oxygen-quenched background data and dividing by the run time.    
Further, any event with more than one distinct pulse in any SiPM during a 12 $\mu s$ event window is removed from the analysis.  

Note that requiring only a single SiPM pulse in any acquisition window should introduce negligible bias to any measurements of dark counts,  
considering that the highest trigger rates from any SiPM-TIA module for any oxygen-quenched run was 46 Hz.  Conservatively assuming this rate is due to all dark noise in a single SiPM, the probability of any two dark pulses overlapping in a 12 microsecond acquisition window is:

\begin{equation}
    R_{DCR} = \tau_1 \cdot \tau_2 \cdot \Delta t = (46 \, Hz) \cdot (46 \, Hz) \cdot (12 \, \mu s) = 0.025 \, Hz
\end{equation}

Compared to the total dark rate, the single pulse requirement cut would remove on the order of $(0.025/46)*100\% = 0.054\%$ of dark pulses, a correction that can be neglected. At the same time, this requirement can strongly suppress SiPM pulses that were induced by scintillation photons.

\begin{figure}[!ht]
    \centering
    \includegraphics[width=0.9\textwidth]{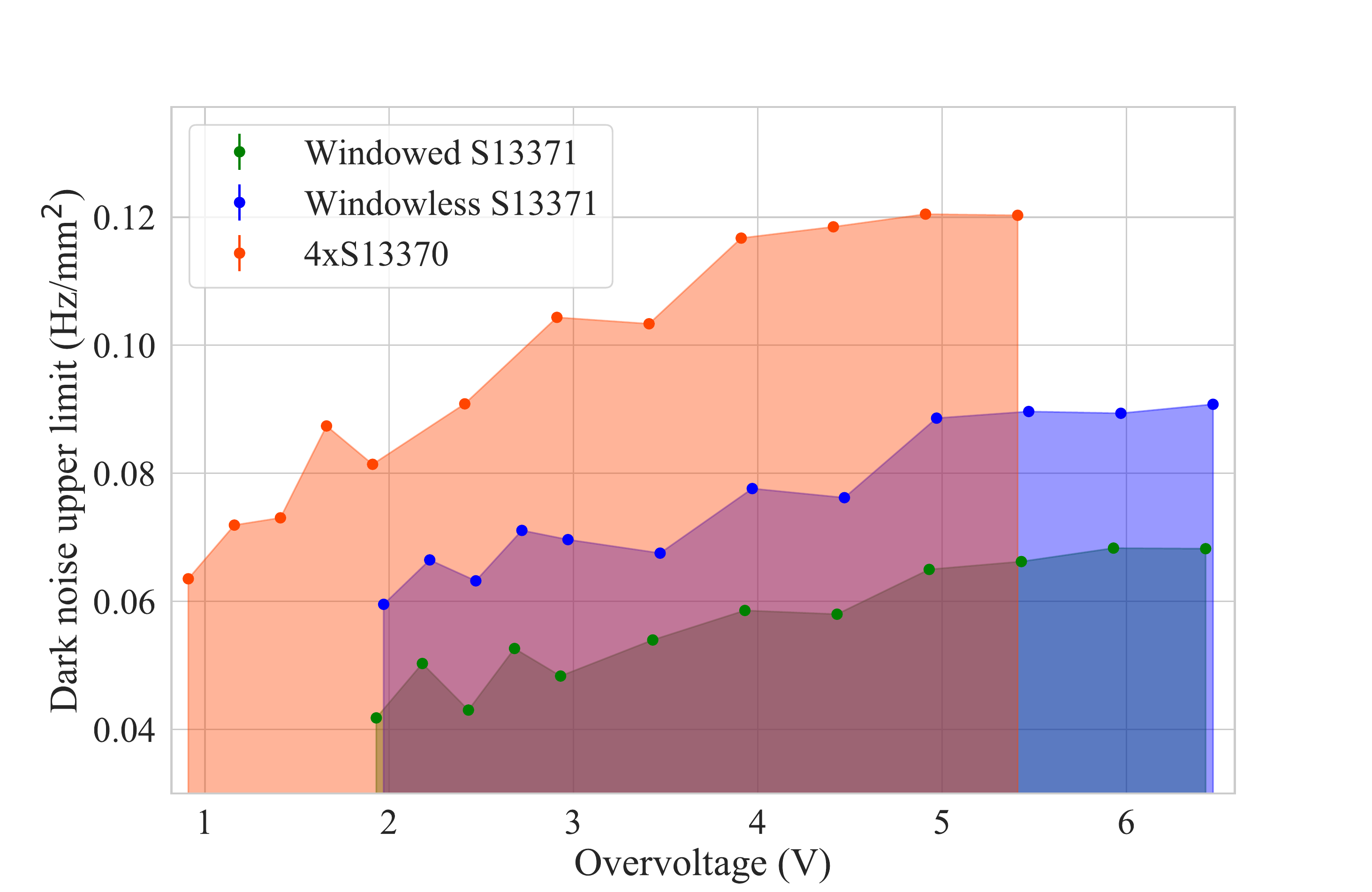}
    \caption{ Upper bound on dark count rate estimates for each Hamamatsu VUV4 SiPM as a function of overvoltage. Error bars reside within the data point itself and are statistical only.  }
    \label{fig:DarkRate}
\end{figure}

The estimated upper bound on dark rates for each SiPM are shown in Figure \ref{fig:DarkRate}.  Each SiPM array has four SiPMs with 6x6 mm$^{2}$ surface area, which is used to normalize the total dark rate estimated in each run.
Notice that the windowless S13371 measures a higher upper bound on the dark noise than that from the windowed S13371; this is likely due to the higher residual scintillation photon contamination expected when no quartz window blocks the SiPM's photosensitive area from the liquid argon volume.

\subsection{Crosstalk and afterpulsing measurements}\label{subsec:CTAP}
There are two intrinsic sources of noise that can shift the mean charge measured from a single photoinduced microcell avalanche. The first source originates from photons produced in a single microcell avalanche traveling to neighboring microcells and producing a second avalanche, known as crosstalk. Due to the shared readout for all microcells within one SiPM unit, crosstalk pulses are always observed as a single pulse with a larger amplification gain. Crosstalk occurs on the nanosecond to tens of nanoseconds scale, and manifests as one or more additional photoelectron charge signals.  
The second source comes from captured and released electrons in the primary avalanche on impurities in the microcell semiconductor, producing a second avalanche in the partially quenched microcell, known as afterpulsing. Afterpulsing predominantly occurs on the tens of nanoseconds scale, but can occur up to a microsecond or two following the prompt avalanche, and has fractional charge and amplitude relative to a single photoelectron \cite{FBK2017,nEXO2018}.  

\begin{figure}[!ht]
    \centering
    \includegraphics[width=0.9\textwidth]{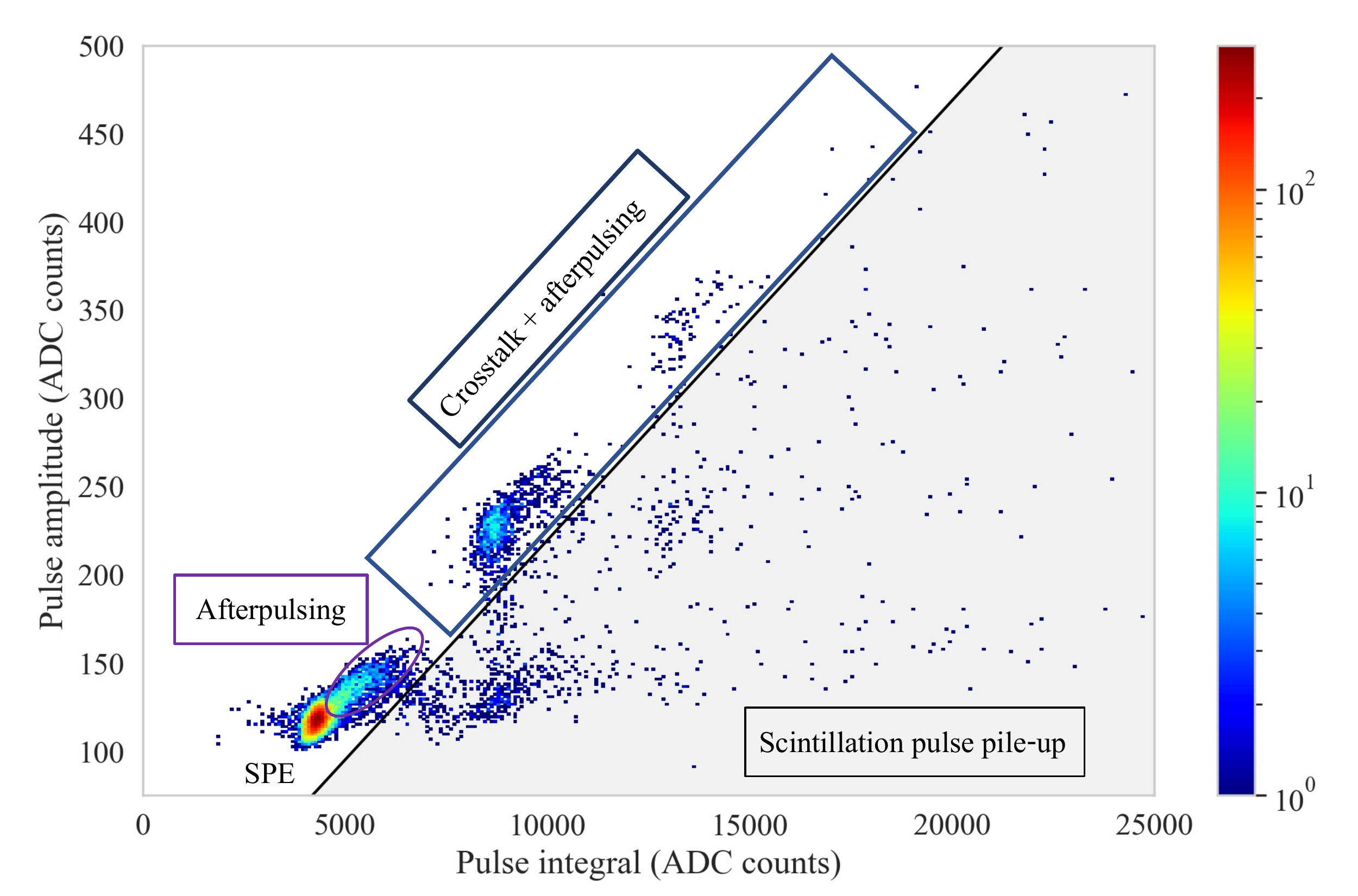}
    \caption{Peak vs. amplitude distribution for pulses in the late window ($>7 \, \mu$s after trigger) of Am-241 calibration runs.  Data is shown for the windowless S13371 SiPM-TIA module at 3.97 V overvoltage.  Relevant pulse noise populations are circled or boxed.  The shaded region below the black line indicates where scintillation pulse pile-up occurs.} 
    \label{fig:APCTExamples}
\end{figure}

The manifestation of crosstalk and afterpulsing in SiPM data taken is shown in Figure \ref{fig:APCTExamples}.  In this analysis we use the photoelectron pulses detected 7 $\mu \mathrm{s}$ after the start of the Am-241-induced argon scintillation prior to oxygen quenching. Given that only tens of photoelectrons are detected in each SiPM for a typical Am-241 event and the rapid exponential decay of the triplet, the probability of more than one photon hitting the SiPM at the same time is estimated to be negligible after 7 $\mu$s. Afterpulsing introduces a higher pulse integral/amplitude tail to the SPE pulse distribution, while crosstalk produces higher photoelectron populations in discrete SPE amplitude/integral intervals. 

The mean charge shift due to afterpulsing in a single photoelectron pulse is estimated in oxygen-quenched data by comparing the single photoelectron fits from section \ref{subsec:Gain} to the total statistical mean of pulses prior to the start of the double photoelectron distribution.  Specifically, the charge shift is calculated as:

\begin{equation}\label{eqn:afterpulsing}
    \Delta Q_{AP}(V) = \frac{\mu_{Q}[0.75 \, PE, 1.75 PE](V) }{\mu_{SPE,fit}(V) } - 1
\end{equation}

The primary systematic uncertainty in this calculation results from the selected pulse area window of [0.75,1.75] PE.  Specifically, the choice of where to define the end of SPE + afterpulses, and equivalently the beginning of double photoelectron pulses, can affect the afterpulsing estimate.  A systematic uncertainty is propagated into the afterpulsing estimates by varying the selected analysis window by 10\% in either direction including the difference in the total uncertainty.  A 10\% variation covers the broadest range between the end of the afterpulsing tail and the start of the second photoelectron distribution without clearly removing the afterpulsing tail or including the second photoelectron peak. The uncertainty of the mean SPE fit is also propagated into the total afterpulsing uncertainty estimate along with the window selection systematic.

\begin{figure}[!ht]
\centering
\begin{minipage}{.497\textwidth}
  \centering
  \includegraphics[width=1.0\linewidth]{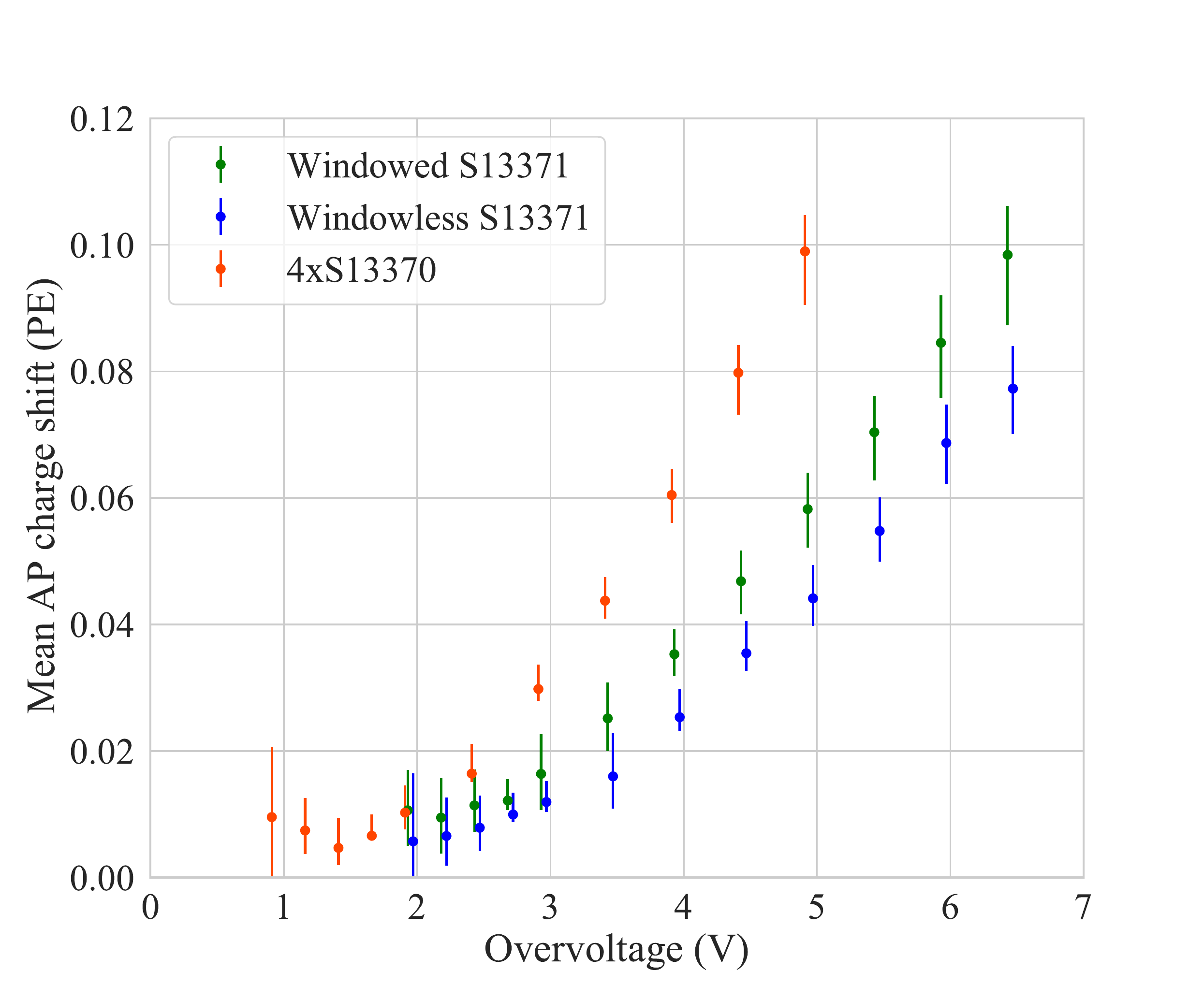}
\end{minipage}
\begin{minipage}{.497\textwidth}
  \centering
  \includegraphics[width=1.0\linewidth]{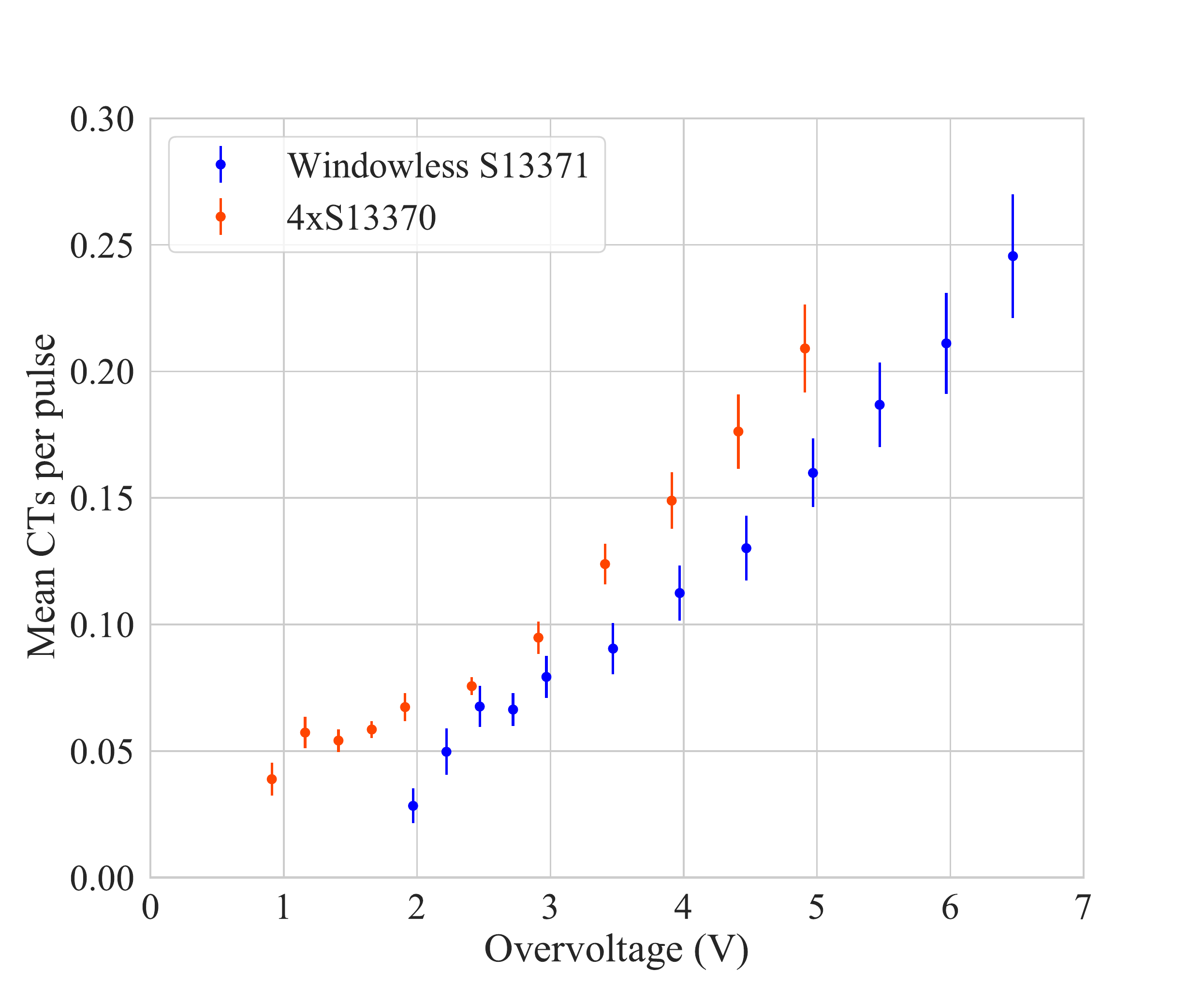}
\end{minipage}%
\caption{Left: Charge shift introduced to the SPE mean due to afterpulsing as a function of overvoltage.  All SiPMs demonstrate an exponential-like increase in afterpulsing as a function of overvoltage. Right: Mean number of crosstalk pulses per single photoelectron as a function of overvoltage for the S13371 and S13370 VUV4 SiPMs.  Both SiPMs demonstrate an exponential-like increase in crosstalk probability as the overvoltage increases.}
\label{fig:CTAP}
\end{figure}

The mean charge contributed to single photoelectron pulses from afterpulsing for each SiPM is shown in Figure \ref{fig:CTAP} (left). An exponential increase in the afterpulsing probability and 
charge shift is observed, in agreement with other measurements in the literature \cite{nEXOSiPM,MarentiniThesis}.  The S13370 units demonstrate a higher afterpulsing probability as a function of overvoltage than both S13371 units.   
The two S13371 units also demonstrate different afterpulsing probabilities, suggesting afterpulsing rates can vary between SiPMs of the same model and that each SiPM package should be characterized individually.

The mean number of crosstalk pulses per single photoelectron is calculated using SiPM pulses in the late tail of Am-241 events with a similar technique described in \cite{nEXOSiPM}.  Modeling the total pulse area for SPEs as composed of afterpulsing from the primary photoelectron and crosstalk pulses and their respective afterpulses, the mean number of crosstalk pulses per SPE is calculated using

\begin{equation}
\bar{N}_{CT}(V) = \frac{1}{N}\sum_{i}^{N}\left[\frac{A_{i}}{1+\Delta Q_{AP}(V)} - 1\right]
\end{equation}
where $N$ is the number of acquisitions, $A_{i}$ is the pulse area of pulse $i$ in units of photoelectrons, and $\Delta Q_{AP}(V)$ is the average charge shift due to afterpulsing in units of photoelectrons.  

There are two main uncertainties in the crosstalk estimate.  The first is the uncertainty of the charge shift due to afterpulsing, as described by equation \ref{eqn:afterpulsing}, while the second is misclassification of single photoelectron pile-up in the Am-241 tail as crosstalk.  Scintillation pile-up contamination is mitigated with a linear cut in pulse integral vs. amplitude space, which rejects pulses with too low of an amplitude-to-integral ratio (see the scintillation contamination population in Figure \ref{fig:APCTExamples}).  
The scintillation pile-up cut's slope is defined by the line between the primary SPE peak and the 2PE peak with double the SPE area/amplitude.  The y-intercept is defined as negative one-fourth the SPE amplitude, which accepts all of the SPE distribution while mitigating scintillation pile-up contamination.  This line cut's y-intercept is varied up and down by 10\% of the SPE amplitude to quantify an uncertainty associated with applying the cut, covering the pulses located between the primary SPE distribution and the valley between the SPE and 2PE scintillation pile-up peak.
Both uncertainties are mitigated by the clear separation between different photoelectron peaks in the pulse amplitude and area distribution provided by the amplification provided by the TIA boards. 

The mean crosstalk pulse count per SPE vs. overvoltage for the windowless S13371 and 4xS13370 arrays are shown in Figure \ref{fig:CTAP} (right). An increase in the crosstalk probability with increased overvoltage is observed for both SiPM models.  The ratio of the 4x13370 unit's crosstalk estimate with that of the S13371 at 4 V overvoltage is in agreement with that quoted by Hamamatsu at room-temperature within uncertainties (see Table \ref{SiPMPropertyTable}).  The crosstalk cannot be estimated for the windowed S13371 using this method, because the quartz window blocks virtually all triplet argon scintillation light and no photoelectron pulses were available in the late tail.

\subsection{SiPM Photon Detection Efficiency (PDE)}

With the basic SiPM characteristics evaluated, we are able to calculate the photon detection efficiency (PDE) of the windowless SiPMs for VUV argon scintillation light, which is defined as the probability that a photon incident on a SiPM's sensitive face will convert into a photoelectron in the device.  The PDE of a SiPM is defined as:
\begin{equation}
    PDE\left(\lambda, V\right) = \nu \left(\lambda\right) \eta \left(V\right) F
\end{equation}
where $\nu$ is the quantum efficiency of the silicon in the SiPM microcells, $\eta$ is the avalanche initiation probability, and $F$ is the SiPM's fill factor \cite{SensLIntro}.  

In this work, the PDE of the windowless S13371 and the 4xS13370 SiPMs is estimated by comparing the measured Am-241 spectrum in each SiPM module before oxygen quenching to that simulated with a Geant4 optical model. The SiPM cells, PCBs, mounting cell, and detector can
geometries are all included in the model, with all surfaces treated as non-reflective except for the SiPM photosensitive areas.  The sensitive SiPM surfaces are assigned a specular reflectivity of 24\%, determined by the mean photon incident angle on the SiPMs in the simulation and reflectivity values reported in \cite{nEXO2019jhg}.  
In the simulation, the Am-241 source is modeled as a point source emitting 59.54 keV gammas at the position outside of the detector can used during calibration, and scintillation photons of 128 nm were generated inside the liquid argon where the gamma energy was deposited.
The absolute light yield of liquid argon used in the simulation is set as
the mean absolute light yield of all data points reported between 30-200 keV in \cite{LArLightYield},
corresponding to a value of $41.1\pm1.2$ photons per keV energy deposited.  
So on average, 2447 primary scintillation photons were generated for every Am-241 gamma ray that deposited full energy in the active liquid argon volume.

\begin{figure}[!ht]
    \centering
    \includegraphics[width=0.95\textwidth]{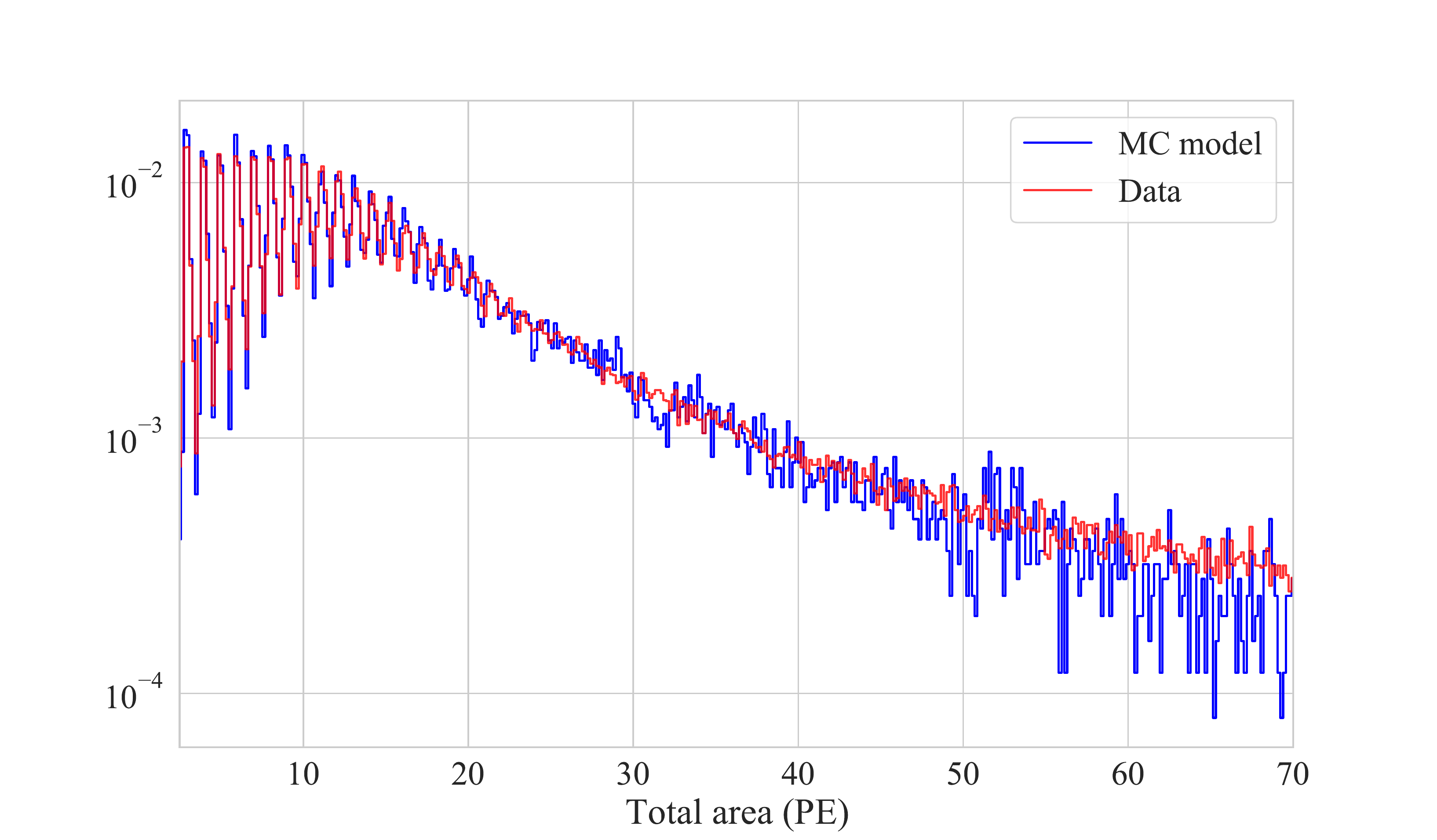}
    \caption{ Comparison of S13371 pulse area distribution at 3.97 V overvoltage collected in an Am-241 calibration run with the model developed using Monte Carlo sampling at PDE=0.164 on simulation data. The model captures the shape in data well, including the individual peaks due to the SiPM's excellent SPE charge resolution.}
    \label{fig:MeanFits}
\end{figure}

Once the expected number of photons incident on the SiPMs for each Am-241 decay event is obtained from the simulation, a binomial distribution with the PDE as a free parameter is used to estimate the number of photoelectrons produced in the SiPMs. Next the single photoelectron charge distributions measured in the tail of Am-241 data for each SiPM module at each overvoltage value (as explained in Sec.~\ref{subsec:CTAP}) is used to simulate the SiPM pulse area detected in this event. By minimizing the chi-square difference between the simulated and observed energy spectra, the SiPM PDE as a function of overvoltage can be derived. An example of the best-fit model developed using simulation and SPE sampling compared to Am-241 data measured by the S13371 module at 3.97 V overvoltage is shown Figure \ref{fig:MeanFits}.

Uncertainties in the sampled SPE distribution due to the scintillation pile-up cut applied to Am-241 data are quantified by varying the cut as described in \ref{subsec:CTAP} and refitting the PDE at each overvoltage.  The difference of the fit values are taken as an uncertainty on the final PDE estimate.   
For SiPM data taken at higher overvoltage values in the SiPM-TIA modules, the prompt Am-241 scintillation tended to saturate the TIAs.  To quantify an uncertainty for this effect, the ratio of charge measured in the late pulse window of [3,9] $\mu \mathrm{s}$ to the full window of [-1,9] $\mu \mathrm{s}$ was compared across all overvoltages in Am-241 calibration data for each SiPM-TIA module.  Deviations in this ratio should be predominantly due to saturation in the prompt peak.  The variation in this ratio indicates that the effect would produce a variation in Am-241 pulse areas of, at most, $\pm 2.5 \%$ in the S13371 module and $\pm 3.5 \%$ in the S13370 module. These uncertainties are propagated into the final PDE estimates.

\begin{figure}[!ht]
    \centering
    \includegraphics[width=0.9\textwidth]{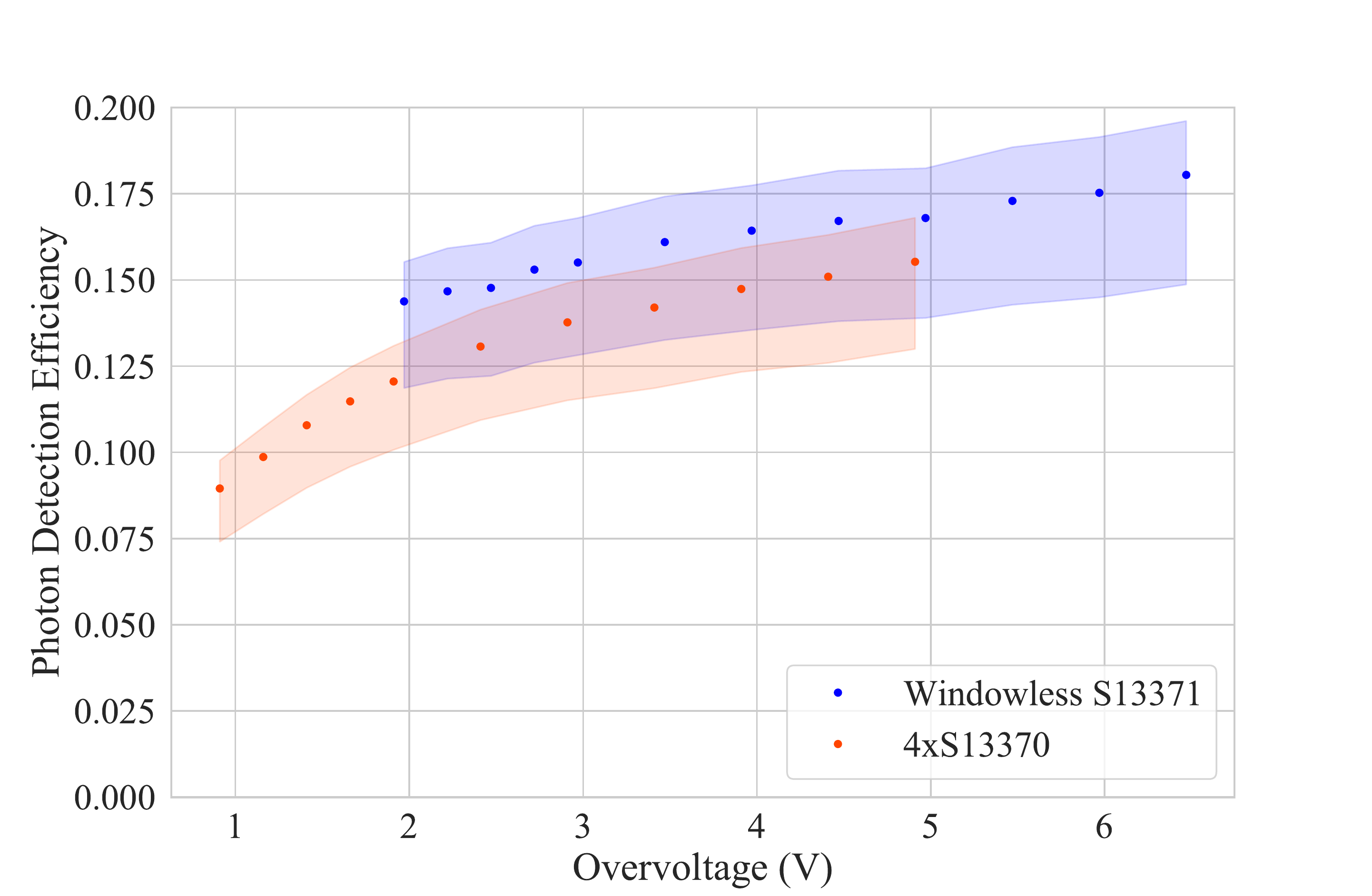}
    \caption{ Photon detection efficiency estimate as a function of overvoltage for the windowless S13371 and 4xS13370 units. The PDE for both models increases with overvoltage, with the PDE's slope gradually decreasing as the overvoltage increases.}
    \label{fig:PDE}
\end{figure}

Additional systematic uncertainty in this calculation may arise from the reflectivity of the Hamamatsu VUV4 SiPM surfaces for 128 nm argon scintillation light. The reflectivity of SiPM surfaces can vary with incident angle, peaking at reflectivity values of up to 
31\% for the VUV4 SiPM surfaces as reported in \cite{nEXO2019jhg}. 
This uncertainty was evaluated by rerunning the simulation with 31\% reflectivity and 0\% reflectivity assigned to the SiPM surfaces and calculating the PDE difference from the aforementioned method. The difference is the largest systematic uncertainty in this work, of order 5\% in the larger PDE direction and 10\% in the smaller PDE direction.   By assuming no reflectivity for 128 nm light from other detector components, our calculated PDE results should be treated as being conservative.

The PDE estimate for the S13371-6050CQ-02 and S13370-6075CQ are shown in Figure \ref{fig:PDE}.  The PDE for both SiPMs is heavily dependent on overvoltage below 4V applied, and drops off rapidly in the 4xS13370 below 2V overvoltage.  This behavior is in agreement with past measurements made using light at xenon scintillation wavelengths \cite{nEXOSiPM}.  
An alternative PDE estimation method of comparing the peak positions in the observed and simulated Am-241 spectra yielded consistent results after corrections for afterpulsing and crosstalk effects were applied.
At 3.97 V overvoltage, the S13371-6050CQ-02's measured PDE is  $17.2^{+1.6}_{-3.0}$\%.  At 3.91 V overvoltage, the S13370-6075CN's measured PDE is $14.7^{+1.1}_{-2.4}$\%.  Both values are consistent with that quoted by Hamamatsu at 4 V overvoltage operation \cite{HamamatsuVUV4Brochure}.   
We note that a similar work using earlier generations of Hamamatsu SiPMs reported VUV PDE values higher than those quoted by Hamamatsu~\cite{Igarashi2016_VUV3}; this work uses additional precautions to mitigate and quantify undesired contamination from longer-wavelength photon backgrounds, and is designed to measure the PDE only for 128 nm argon scintillation photons.

\section{Conclusions}

We deployed three Hamamatsu VUV4 SiPMs, including one S13371 unit with a quartz window, one S13371 unit without a window, and another module containing four windowless S13370 units, in a liquid argon detector. Each SiPM module was equipped with a Texas Instruments model LMH6629 operational amplifier to improve the single-photoelectron signal quality. By carefully examining the SiPM signals under different experimental conditions and by comparing performance of SiPM modules with different wavelength sensitivities, we unambiguously confirmed the sensitivity of Hamamatsu VUV4 SiPMs for 128 nm liquid argon scintillation light. We further characterized the gain, dark count rate, crosstalk and afterpulsing probabilities of these devices in a liquid argon environment. Using an Am-241 radiation source to produce argon scintillation photons of a known intensity, we obtained a photon detection efficiency of $14.7^{+1.1}_{-2.4}$\% and $17.2^{+1.6}_{-3.0}$\% for the S13370-6075CN and S13371-6050CQ-02 VUV4 SiPMs at 128nm wavelength and 4 V overvoltage operation. 


\section*{Acknowledgments}

This work was performed under the auspices of the U.S. Department of Energy by Lawrence Livermore National Laboratory (LLNL) under Contract DE-AC52-07NA27344 and was supported by the LLNL-LDRD Program under Project No. 20-SI-003.
It is also supported by the U.S. Department of Energy (DOE) Office of Science, Office of High Energy Physics under Work Proposal Number SCW1676 awarded to LLNL. 
J. Kingston is supported by the DOE/NNSA under Award Number DE-NA0000979 and DE-NA0003996 through the Nuclear Science and Security Consortium. 
I. Jovanovic is partially supported by the Department of Energy National Nuclear Security Administration, Consortium for Monitoring, Technology, and Verification (DE-NA0003920)

We thank Nathan Eric Robertson and Phillip Hamilton from LLNL for their technical support on fabricating parts for the liquid argon detector. 

LLNL IM release number: LLNL-JRNL-831432

\end{document}